# Accessible, All-Polymer Metasurfaces: Low Effort, High Quality Factor

*Michael Hirler, Alexander A. Antonov, Enrico Baù, Andreas Aigner, Connor Heimig, Haiyang Hu and Andreas Tittl\**

Chair in Hybrid Nanosystems, Nanoinstitute Munich, Faculty of Physics, Ludwig-Maximilians-Universität, Munich, Germany

\* Email: Andreas.Tittl@physik.uni-muenchen.de

KEYWORDS: polymers, metasurfaces, bound states in the continuum, nanofabrication, low-index photonics.

**ABSTRACT**

Optical metasurfaces supporting resonances with high quality factors offer an outstanding platform for applications such as non-linear optics, light guiding, lasing, sensing, light-matter coupling, and quantum optics. However, 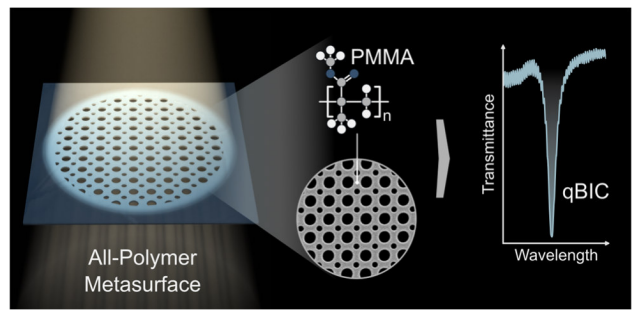 their experimental realization typically demands elaborate multi-step procedures such as metal or dielectric deposition, lift-off, and reactive ion etching. As a consequence, accessibility, large-scale production and sustainability are constrained by reliance on cost-, time- and labor-intensive facilities. We overcome this fabrication hurdle by repurposing polymethyl methacrylate—which is usually employed as a temporary resist—as the resonator material, thereby eliminating all steps except for spin-coating, exposure and development. Because the low refractive index of the polymer limits effective mode formation, we present a bilayer recipe that enables the convenient fabrication of a freestanding membrane to maximize the index contrast with its surroundings. Since etching induced defects are circumvented, the membrane features high quality nanopatterns. We further examine the suspended membrane with scanning electron microscopy and extract its position-dependent spring constant and pretension with nanoindentation experiments applied by the tip of an atomic force microscope. Our all-polymer metasurface hosting Bound States in the Continuum experimentally delivers high quality factors (up to 523) at visible and near infrared wavelengths, despite the low refractive index of the polymer, and enables straightforward geometry-based tuning of both linewidth and resonance position. We envision this methodology to lay the groundwork for accessible, high performance metasurfaces with unique use cases such as material blending, angled writing and mechanically based resonance tuning.

Optical metasurfaces are ultrathin flat photonic structures that owe decisive parts of their optical properties to their artificial sub-wavelength features, enabling efficient manipulation of light at the nanoscale beyond the capabilities of conventional bulk materials.[1–3]. Inspired by findings such as the generalization of Snell's law by Yin *et al,.*[4] metasurfaces have reshaped the landscape of flat optics, achieving landmarks in the level of control over phase,[4–7] polarization,[6,8] amplitude[5,9,10] and dispersion[11] of incident waves with sub-wavelength resolution. In particular, they can be engineered to strongly suppress scattering losses and support resonances with high quality ($Q$) factors through various physical mechanism.[12–14] One of these mechanisms has recently attracted increasing attention: metasurfaces supporting Bound States in the continuum (BICs) offer a high degree of control over radiative losses.[15–17] A true BIC is a perfectly confined mode without any radiation, yet residing the continuum of radiating modes. By perturbing this ideal lossless mode, a resonance known as quasi-BIC (qBIC) becomes accessible in the far field. QBICs have enabled a wide range of applications including non-linear optics,[18] light guiding,[19] lasing,[20] sensing,[21–24] light-matter coupling[25–28] and quantum optics.[29]

Although qBICs have been demonstrated in both metallic[30–32] and dielectric[21,33,34] structures, the significant resistive losses in plasmonic resonators have led to the widespread use of high-index and low-loss dielectrics such as silicon and germanium.[35,36] However, restricted material diversity limits the scope of applications, missing out on the potential other materials might offer: Not only are low-loss, high-index materials rare at visible wavelengths,[37,38] but patterning them into high-quality nanostructures demands a sophisticated infrastructure that is cost-, time- and labor-intensive, ultimately limiting large-scale commercialization.[39] Moreover, etching-based procedures are prone to fabrication errors such as geometric perturbations[40–42] and material redeposition.[43,44] At the same time, the complex fabrication demands substantial energy and material consumption, along with intensive use of gases and chemicals, resulting in a significant ecological footprint and safety risk. Gases like chlorine and chemicals like hydrofluoric acid, for instance, require careful disposal and pose the risk of leaks or hazardous accidents.[45,46]

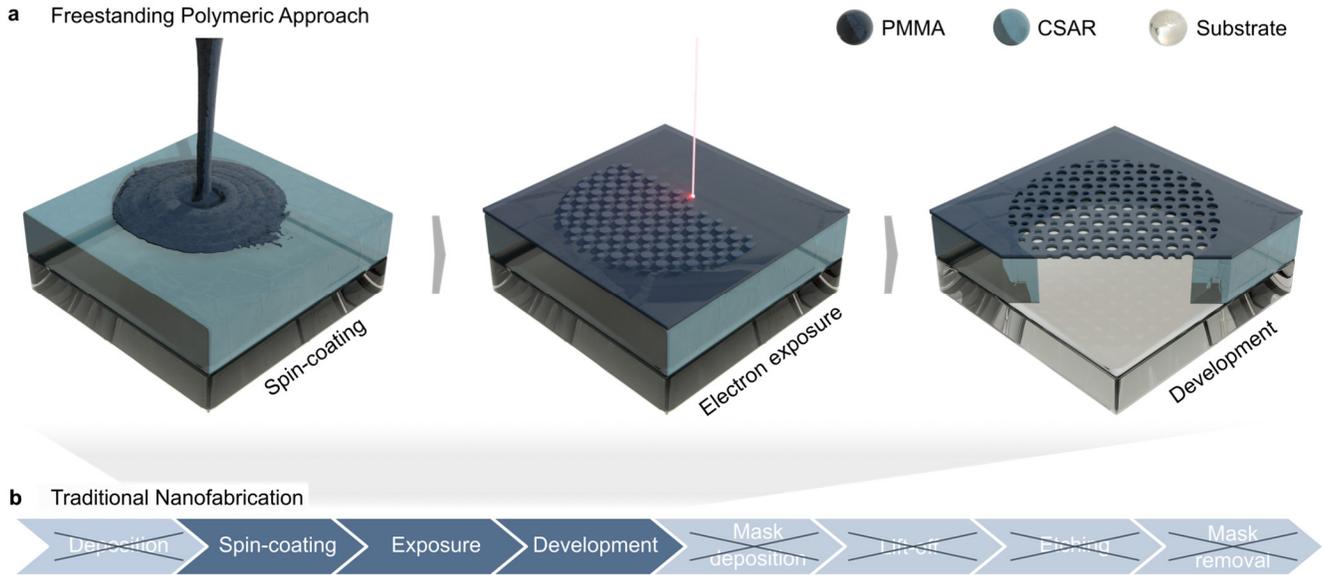

**Figure 1**. Simplified Fabrication Workflow. (a) Schematic illustration of the nanofabrication workflow for an all-polymeric metasurface, solely consisting of spin-coating, exposure and development. This greatly simplifies the traditional fabrication process as shown in (b).

Polymers are safe, affordable and easy to process.[47] In fact, they already play an integral role in state-of-the-art lithography processes.[48,49] In the form of resists, they are used to transfer patterns onto the target material. Typically, they serve a temporary purpose and are removed from the final device during the lift-off process. A conventional fabrication workflow of metasurfaces is schematically illustrated in Figure 1b: the deposition of a metal or dielectric is followed by spin-coating of the resist, exposure with photons or electrons, resist development, deposition of the etching mask, lift-off, etching and the removal of the mask. Among common polymers, polymethyl methacrylate (PMMA) is arguably the most available resist. It is well-established, extensively studied and has been effectively combined with other resists for multi-layer precedures.[50,51]

However, the low refractive index (RI) of polymers hampers their utility as integral constituents of resonant metasurfaces, since the dissipation of optical power into the substrate inhibits the effective formation of confined modes. Metallic substrates such as gold or silver have been proposed to prevent mode leakage by interaction with surface plasmons.[52,53] For instance, Kulkarni *et al.* demonstrated qBICs with $Q$ factors up to 305 at visible wavelengths emerging from a polymer resist film placed on a silver substrate.[54] But the ohmic losses in the substrate eventually limit the potential of BICs, restricting the diverging nature of their $Q$ factors.[53] In addition, the use of metallic mirrors restricts the device to operate in reflection, whereas many applications including cascaded metasurfaces and device-integration[55,56] favor transmission. A suspended polymeric photonic crystal nanocavity presented by Martiradonna *et al.* circumvented substrate induced losses as well as restriction to reflection mode altogether, but came at the cost of reversion to elaborate fabrication procedures involving hazardous chemicals.[57] To date, a simple and cost-effective fabrication technique for a suspended, nanopatterned polymer membrane is still pending.

In this work, we realize an all-polymer metasurface based on qBICs. Our approach drastically simplifies the nanofabrication process, solely relying on spin-coating, exposure and development (Figure 1a). The result is a freestanding PMMA membrane whose photonic behavior is independent of the parasitic diffraction and losses associated with the substrate. We numerically simulate a $C_4$-symmetric hole design hosting both, an electric and a magnetic BIC. Then, we outline our bilayer fabrication procedure before we verify the formation of high quality nanopatterns in suspended polymer films with scanning electron microscopy (SEM). Nanoindentation experiments applied by the tip of an atomic force microscope (AFM) further retrieve the membrane's position-dependent spring constant and pretension. Finally, we experimentally demonstrate that the patterned polymer films support qBICs at visible and near infrared wavelengths with $Q$ factors as high as 523 along with geometry driven tunability of the resonance's position and linewidth.

**RESULTS AND DISCUSSION**

**Numerical Modelling of a Free-Standing Metasurface with BZF-BIC.** As basis for our platform, we utilized a design consisting of periodic hole arrays in a freestanding PMMA membrane, where the quadratic unit cell encloses a void at its center (Figure 2a). Assuming the absence of internal losses and for normal incidence, the bound state is non-radiative with an infinitely high $Q$ factor as long as the holes are identical in size. Such a perfectly confined state cannot couple to free space light. In order to access the otherwise dark BIC state, we perform Brillion zone folding (BZF) in momentum space by periodically varying the radii of adjacent holes. The perturbation parameter $\alpha = \frac{r_1 - r_2}{r_1} = \frac{\Delta r}{r_1}$ quantifies this variation. This new configuration extends the unit cell by a factor $\sqrt{2}$ and introduces a radiative leakage channel, converting the true BIC into qBICs with high but finite $Q$ factors. In the far field transmittance spectrum (Figure 2b), this manifests in the emergence of two BZF induced qBICs; we distinguish them by their in-plane field distribution as an electric and a magnetic mode. Their linewidth and resonance position depend on $\alpha$, which exemplarily demonstrates the large degree of control BIC based structures allow to exert over the radiative loss channel.[58] The resonance wavelength can be controlled by other geometrical parameters[21,25,28,38]—here, the resonance is designed to reside at visible wavelengths, with a membrane thickness $t$ of 300 nm and unit cell periodicities $p$ of 410 nm (unperturbed design) and 580 nm (perturbed design). The holes are chosen such that the combined area $A$ of two holes is conserved upon changes in $\alpha$, i.e. $A = 2\pi r_0^2$ where $r_0 = 135.83$ nm for the unperturbed and $A = (r_1^2 + r_2^2)$ for the perturbed configuration. It directly follows that $r_1 = \sqrt{A[\pi(1+(1-\alpha)^2)]^{-1}}$ and $r_2 = r_1(1-\alpha)$. The RI of PMMA was obtained from in-house ellipsometry measurements (SI Note 1) and can be approximated as 1.5. Since the magnetic qBIC features very narrow linewidths that are difficult to access experimentally, we focus on the electric mode throughout this work. Figure 2c indicates the electric near field enhancement of the perturbed configuration for $\alpha = 0.5$. The fields are efficiently concentrated in the voids (maximum enhancement of $\left|\frac{E}{E_0}\right| \approx 18$, where $E_0$ denotes the amplitude of the incident electric field), which makes the platform a promising candidate for effective near field light-matter coupling.[28] Considering momentum space, we demonstrate that at the Γ-point (*i.e.* $k_x = k_y = 0$), there are two degenerate electric modes, which enables their coupling to the far field. In contrast, non-degenerate modes in a structure with $C_4$ rotational symmetry cannot radiate.[59] Deviations from the Γ-point break $C_4$ rotational symmetry and lift the qBIC's degeneracy, resulting in two electric modes depending on the polarization. We label them mode 1 and mode 2. Interestingly, mode 1 exhibits a pronounced angle-stability in the directions where $k_x = 0$ and $k_y = 0$, respectively (Figure 2d), while the *Q* factor remains roughly constant (Figure 2e). Mode 2, in contrast, is significantly more sensitive along those directions. This implies that, depending on the polarization, the system can be chosen to exhibit either robustness or sensitivity upon changes of the angle of the incident light. Further simulations of the far field polarization unveil the topological nature of the modes. Both vortices show integer winding numbers of the polarization vectors around the vortex centers, corresponding to topological charges of -1, which points at the underlying physics of qBICs. Analogous considerations for the magnetic qBIC can be found in SI Note 2.

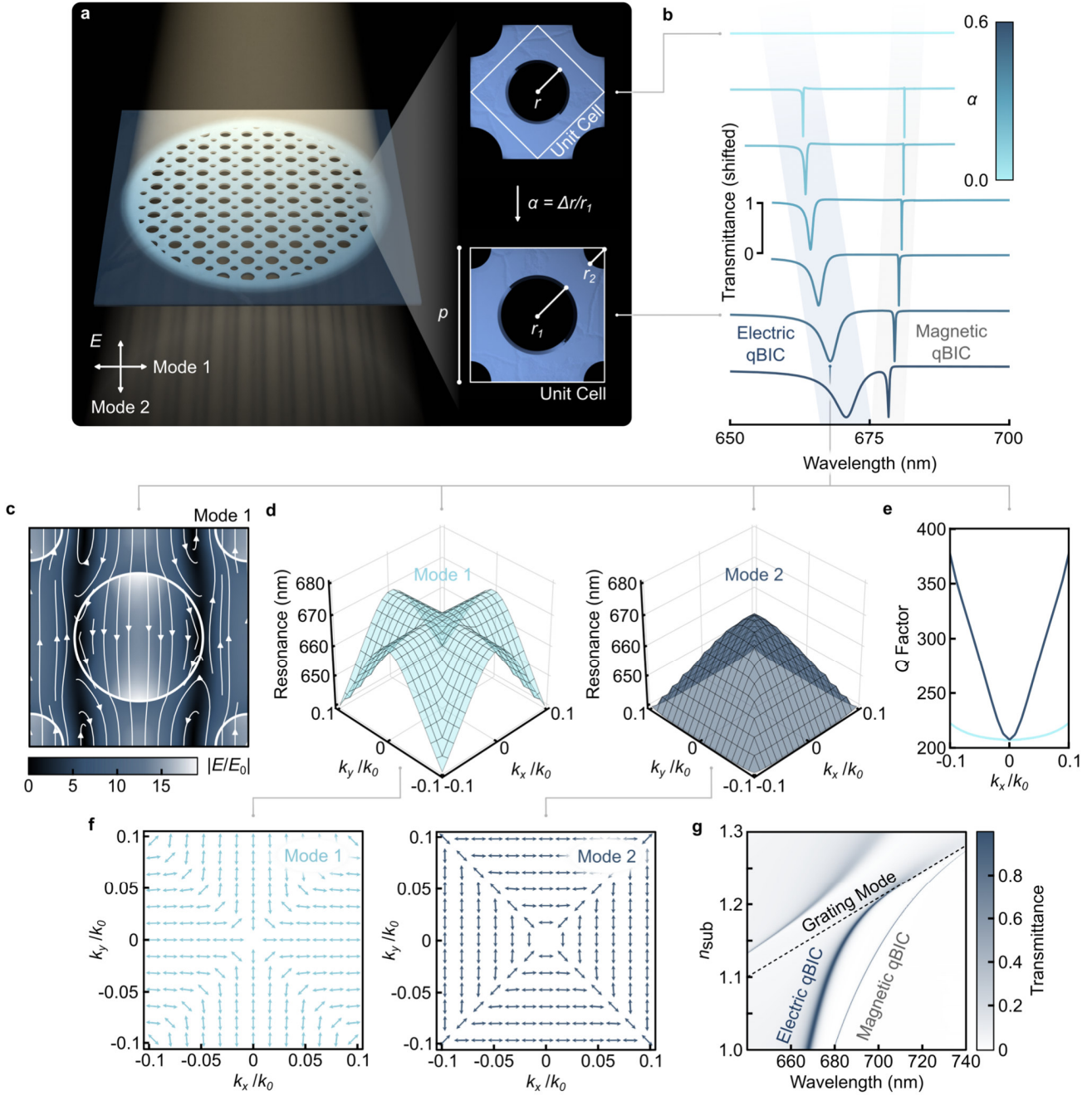

**Figure 2**. Numerical Modelling of a Free-Standing Metasurface with BZF-BIC. (a) Artistic rendering of a freestanding membrane featuring hole arrays that support true (top) and BZF-quasi (bottom) BIC. (b) Transmittance spectra for different perturbations $\alpha$. (c) Electric near field enhancement $|E/E_0|$, resonance position (d) and $Q$ factor (e) in momentum space as well as far field polarization (f) for the two electric modes that are degenerate at the Γ-point. (g) Transmittance spectra depending on the RI of the substrate $n_{sub}$.

Our simulations confirm that bound states in the continuum provide an effective mean to achieve confined modes in a freestanding PMMA membrane in spite of the polymer's low RI of roughly 1.5. However, most commonly, metasurface are fabricated on substrates. Not only does this grant mechanical stability, but also simplify the fabrication procedure. This is where the use of low-index polymers poses a significant hurdle: much like in conventional optical gratings, periodic metasurfaces give rise to discrete diffraction orders. Such grating orders—also known as Rayleigh-Wood anomalies[60,61]—are a consequence of the structure's periodicity $p$ and their wavelength $\lambda$ follows $\lambda = \frac{n_{sub} \cdot p}{m}$ for normal incidence of the light (where $m$ is an integer indicating the diffraction order). To prevent interference and thus suppression of the resonance, spectral proximity between qBIC and first grating mode needs to be avoided. As Figure 2g illustrates, for a patterned PMMA membrane resting on a substrate, even low values of the substrate's index $n_{sub}$ result in such interferences. Due to the low RI of PMMA, this cannot be prevented by geometrical corrections. Consequently, the challenge is to either employ ultra-low index substrates[62,63] or, preferably, to manufacture a freestanding polymer membrane. While there have been demonstrations of freestanding membranes, for instance, consisting of silicon by Adi *et al.*,[28] the potential of polymers, which are easier to process, has not yet been fully harvested.

**Experimental Realization and Examination of Freestanding PMMA Metasurface.** In order to build freestanding membrane without reverting to cumbersome deposition and etching steps, we propose a top-down bilayer resist recipe: We spin-coated a 300 nm thick PMMA film on top of a ca. 1250 nm thick sacrificial layer consisting of CSAR resist. The PMMA was solved in ethyl lactate to prevent mixing with the underlying layer. During the subsequent electron beam exposure, the different sensitivities of the two polymers were exploited. While applied doses ranging from 200 to 250 µC/cm² patterned the nanohole array into the PMMA layer, the underlying CSAR was overexposed instead. This selective exposure allowed the sacrificial CSAR layer to be entirely removed through the porous PMMA film during development, releasing a freestanding membrane (Figure 1a). More details can be found in the methods section. Owing to the use of polymers, neither deposition, etching nor lift-off were required. Reducing the required number of steps also reduces the procedure's proneness to errors.

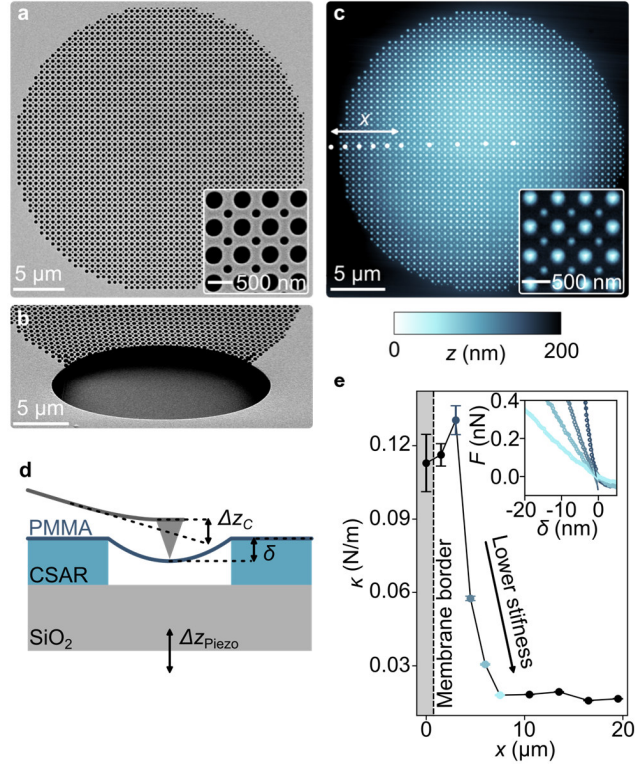

**Figure 3**. Fabricated Membrane under Mechanical Load. SEM images of (a) a fabricated metasurface with $\alpha = 0.5$ and (b) a metasurface featuring a notch to illustrate the suspension (viewing angle 45°). (c) Measured topography of the membrane. The white dots indicate the positions of the retrace curves shown in (e). (d) Schematic diagram of the nanoscopic bending test experiment. (e) Membrane spring constant $\kappa(x)$ as function of the distance $x$ from the surrounding bulk material obtained from fitting Eq. (1) to retrace curves. The inset exemplarily shows some of these force-displacement ($F$-$\delta$) curves. The dashed black line delineates the border of the metasurface from the surrounding bulk polymer. Error bars show the standard deviation of $\kappa$ fitted to each retrace curve.

We examine the patterned and suspended membrane in two ways: first, Figure 3a shows SEM images of the final device. The metasurface's circumference was designed to be circular rather than rectangular to homogenize the strain profile, as sharp corners were found to promote fractures in the polymer film (SI Note 3). The diameter measures 30 µm. The film in Figure 3b additionally features a notch to verify the suspension of the metasurface. Noteworthy, the hanging PMMA shows little to no bending despite its thickness of only 300 nm, while the holes exhibit high quality with little imperfections. Since the polymeric approach circumvents any pattern transfer, the PMMA was not damaged by etching. Consequently, the precision of the final structure almost solely relies on the high-resolution lithography step. Second, to study the mechanical properties of our all-polymer metasurface, AFM measurements were performed, both in tapping and contact mode in order to extract the spring constants and pretensions of the free-standing membrane, respectively. Unlike in Figure 3a, the image taken in tapping mode (Figure 3c) shows a depression at the location of the metasurface relative to its surroundings, suggesting that the mechanical load applied by the tip bends the elastic membrane downwards. Furthermore, we conducted contact mode retrace curves at several positions across the metasurface (dots in Figure 3c) with an initial force set to around 0.6 µN. Starting from the border and moving towards the center, each position was measured ten times, resulting in the averaged force-displacement curves exemplarily presented in the inset of Figure 3e (all curves are shown in SI Note 4). The measurements suggest that the membrane stiffness decreases with distance $x$ from the bulk material around the metasurface, indicated by the shallower slopes. To obtain the elastic membrane deformation $\delta$, we subtract the cantilever deflection $\Delta z_c$ from the displacement of the scanning piezotube of the AFM instrument $\Delta z_{\text{piezo}}$ as $\delta = \Delta z_{\text{piezo}} - \Delta z_c$ (Figure 3d). $\Delta z_c$ can be obtained by recording a reference retrace curve on silicon (SI Note 4). To describe the bending of our metasurface, we employ a simplified model of Kirchhoff plate theory that relates the applied force to the deformation of the membrane.[64] Since the metasurface radius $R$ of 15 µm is much larger than its thickness $t$ of 300 nm, we consider the relation between loading force $F$ and displacement $\delta$ in the limit $\frac{t}{R} \ll 1$, given by the following form:[65,66]

$$F = \left[\frac{4\pi E}{3(1-v^2)}\frac{t^3}{R^2} + \pi T\right]\delta \quad (1)$$

where $E$ is Young's modulus, $v$ the Poisson ratio of PMMA and $T$ is the pretension of the fabricated membrane. Since the relationship between $F$ and $\delta$ is linear for this approximation, we fit a slope to the measured retrace curve, which corresponds to the membrane spring constant $\kappa = \frac{\partial F}{\partial \delta}$. As seen in the retrace curves, the extracted spring constant is position-dependent and decreases sharply with increasing distance from the border, ranging from 120 to 20 N/m (Figure 3e). Both, the order of the spring constant's magnitude and its qualitative behavior across the membrane are similar to previous conducted studies on suspended nanomembranes consisting of polymeric layers with gold nanoparticle intralayers.[67,68] We observe that the contribution to the spring constant stems mainly from the pretension $T$, which is remarkably high for such a soft material. We attribute these values to fabrication processes such as baking, which could induce stress on the PMMA before suspension.

**Optical Characterization of the Metasurface.** Experimental spectra were recorded in transmittance. The measurements in Figure 4a show good agreement with the simulations (Figure 2b): while in the unperturbed case with identical hole radii ($\alpha = 0$) the mode is completely decoupled from the radiative continuum and thus absent the spectrum, the radius perturbations ($\alpha > 0$) open radiative leakage channels and the electric mode resonance emerges. The highest $Q$ factor measured was 523 (Figure 4b) as determined by fitting a Fano resonance. We attribute deviations from the simulations to non-ideal illumination conditions as well as finite size effects.[69] Besides control over the linewidth, we further demonstrate tunability of the resonance position by scaling the unit cell's lateral dimensions by a scaling factor $S$. Figure 4c shows the corresponding transmittance spectra: as *S* increased from 0.8 to 1.3, the electric qBIC responds by shifting from 551 to 838 nm; that is, from visible to near infrared wavelengths. Note that the spectra were individually normalized such that the minimum and maximum transmittance equals 0 and 1, respectively (raw data are reported in SI Note 5). The geometry driven tunability of the qBIC enables high adaptability to application-specific demands.

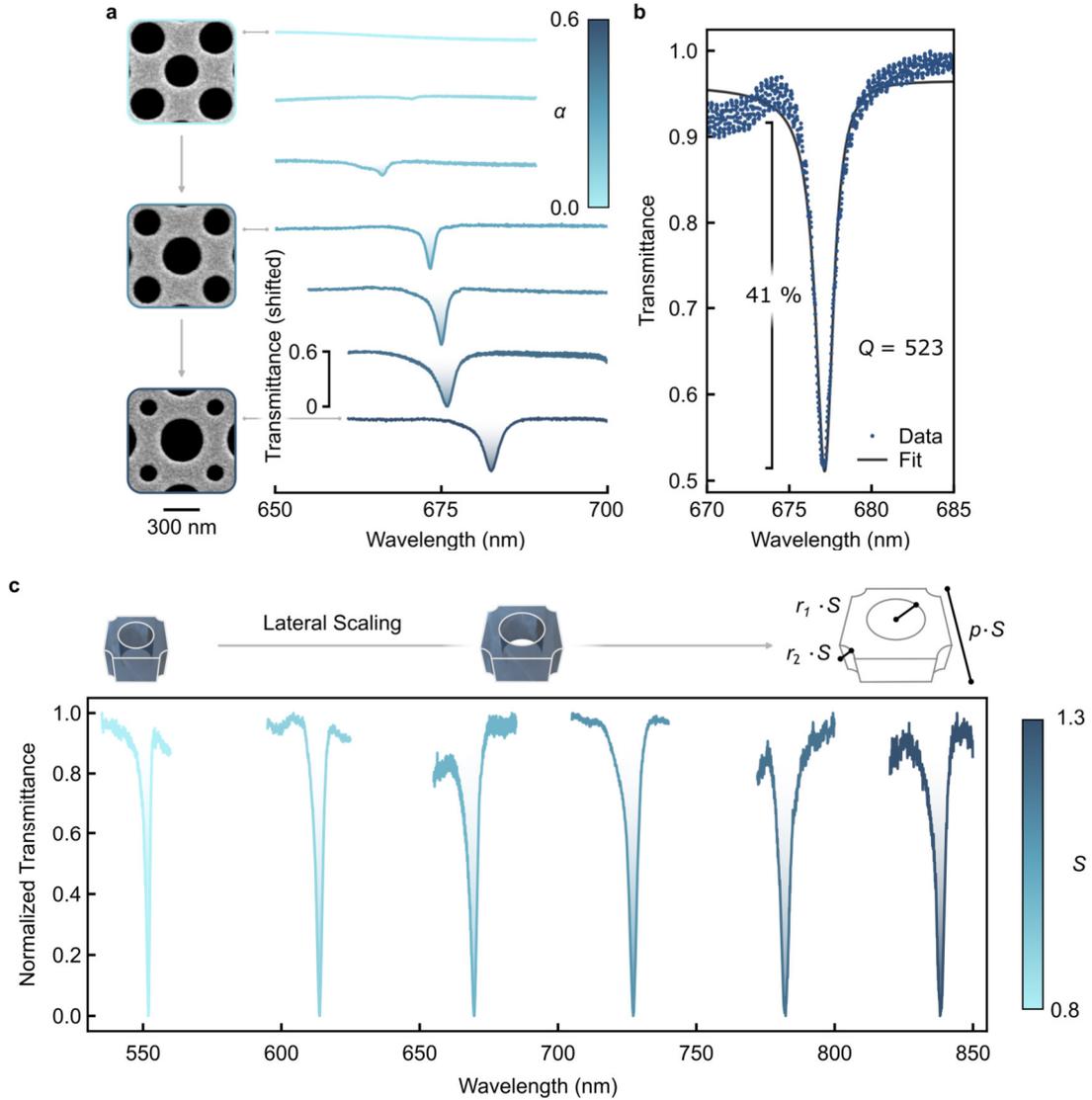

**Figure 4**. Experimental Spectra. (a) SEM close ups of various perturbation parameters $\alpha$ along with corresponding transmittance spectra of the electric mode. (b) Resonance with highest measured $Q$ factor of 523 and amplitude of more than 40 %. (c) Transmittance spectra normalized from 0 to 1 from the visible to near infrared region obtained from lateral scaling of the unit cell with $\alpha = 0.5$ by a factor $S$.

**CONCLUSION**

In essence, we developed an all-polymer metasurface with geometrically tunable qBICs at visible and near infrared wavelengths with *Q* factors up to 523. Our bilayer fabrication methodology—consisting solely of spin-coating, exposure and development—outperforms conventional fabrication procedures in terms of facility requirements, time and labor costs as well as environmental footprint. At the same time, since etching induced defects are circumvented, it delivers high quality nanopatterns. Utilizing photolithography instead of electron beam lithography could also significantly reduce exposure time, paving the way for high-throughput commercialization. Moreover, we envision extensions of our method to enable unique use cases such as material blending,[70] for instance with emitting particles,[54] and angled electron exposure for chiral photonics. Our suspended membrane metasurface is further able to withstand loads of at least 0.6 µN without rupturing and exhibits spring constants of 20 N/m and higher. Such mechanical stability suggests that our platform could be directly used to build robust, yet inexpensive nanosensors resilient to environmental perturbations such as vibrations and other forms of mechanical stress.[71] The polymer's stability could be further increased by crosslinking. The membrane's flexibility may further offer a promising basis for reversible mechanically tunability of the geometric parameters and thus the optical response, potentially in conjunction with simultaneous strain-engineering of materials such as two-dimensional transition metal dichalcogenides for photoluminescence enhancement.[72] We hope to fortify the foundation for accessible, high performance metasurfaces with a variety of unforeseen applications.

**METHODS**

**Numerical Modelling.** The numerical simulations in Figure 2a-c, 3g and S4a were performed in CST Studio Suite 2023 (Daussault Systèms), a commercial finite element solver. Operating in frequency domain, we assumed periodic boundary conditions in x- and y-directions and configurated adaptive mesh refinement. The RI of PMMA was imported from in-house ellipsometry data (SI Note 1). To solve the eigenstates problem (Figure 2d-e and S2b-c) the electromagnetic waves frequency domain module of COMSOL Multiphysics in 3D mode with the eigenstate solver was used. The tetrahedral spatial mesh for Finite Element Method was automatically generated by COMSOL's physics-controlled preset. Simulations were performed within a rectangular spatial domain containing a single metasurface unit cell with periodic boundary conditions applied to its sides. To calculate the far field polarization (Figure 2f and S2d) of the eigenstates the overlap integrals between the eigenmodes displacement current and plane waves were estimated using a previously developed approach.[73]

**Fabrication of Freestanding Polymer Membranes.** Prior to the applications of the resists, the adhesion promoter SurPass 4000 (MicroResist) was coated (500 rpm for 30 s), onto cleaned fused silica substrates (MicroChemicals), washed off with isopropanol and spin-dried (3000 rpm for 30 s). Next, two layers of CSAR 62 (AR-P 6200.13, Allresist) were successively applied (1000 rpm for 1 min) and baked at 170 °C for 5 min, adding up to a ca. 1250 nm thick sacrificial layer. A 300 nm thick layer of PMMA (AR-P 679-04, Allresist) was then added on top (3600 rpm for 1 min) and baked at 150 °C for 3 min. The conductive polymer Espacer 300Z (Showa Denko K.K) finalized the spin-coating procedure (2000 rpm, 1 min) and prepared the sample for the following electron beam lithography treatment. To this end, an eLINE Plus system (Raith) operating at 20 KV with 15 μm aperture wrote the metasurface geometry into the films with doses ranging from 200 to 250 μC/cm$^2$. Finally, the patterned sample was immersed, first, in a 7:3 isopropanol-water blend for 20 s to develop the PMMA and, second, in cold Amyl Acetate (Supelco) at 6 °C for 10 s to remove the underlying CSAR. The process was eventually stopped in Novec 7100 (Sigma-Aldrich), which evaporates readily without requiring blow-drying. The procedure yielded a patterned, freestanding polymer membrane.

**SEM Imaging.** SEM images were taken with a Gemini device from Zeiss operating at 2-3 KV. Beforehand, a few nm thin palladium-gold film was sputtered on the samples to increase the contrast and reduce charging effects. Afterwards, some images were post-processed by carefully adjusting the brightness and contrast.

**Optical Characterization.** The spectra of the metasurfaces were recorded with a commercial white-light confocal optical microscope (WiTec alpha 300 series, Oxford Instruments) in transmittance mode. Illumination was directed from below with linearized collimated light from a broadband halogen lamp (OSL2, Thorlabs), which was collected by a 20x objective (Zeiss, NA = 0.4) and coupled into a multimode fiber. The collected light was dispersed by a diffraction grating with 600 mm$^{-1}$ groove density (1800 in Figure 4b) and subsequently detected by a silicon CCD sensor. To remove undesired features from the sample or the beam path, we referenced all our measurements with the transmittance spectra of the unpatterned sample. Each individual spectrum was obtained from 10 accumulations, with an integration time of 0.5 s per accumulation.

**Determination of $Q$ Factors.** Temporal Coupled Mode Theory[74,75] was employed to determine the $Q$ factors of the resonances (as in Figure 4b). To this end, we fitted the following equation for the transmittance $T$ to the spectra:

$$T = \left| e^{i\phi} t_o + \frac{\gamma_r}{\gamma_r + \gamma_i + i(\lambda - \lambda_{res})} \right|^2 \qquad (2)$$

where $e^{i\phi} t_o$ describes the background transmission, $\lambda_{res}$ the resonance wavelength and the loss rate $\gamma = \gamma_r + \gamma_i$ includes the radiative ($\gamma_r$) and intrinsic ($\gamma_i$) losses. The $Q$ factor can be subsequently determined as $Q = \frac{\lambda_{res}}{2\gamma}$.

**AFM Experiments.** The AFM measurements were conducted using a commercially available AFM-based scattering scanning near-field optical microscope (NeaSNOM, Attocube Systems), operating both in tapping mode for imaging (Figure 3c) and contact mode for recording retrace curves (Figure 3e). The AFM tip (Arrow-NCPt, NanoWorld) used has a cantilever spring constant of $\kappa_c = 42$ N/m and a resonance frequency of $\Omega \approx 250$ kHz in tapping mode. Integration times were set to around 10 ms per pixel and tapping amplitudes ranged between 70 and 80 nm.

## ASSOCIATED CONTENT

**Supporting Information Available:**

Refractive index data of PMMA; numerical modelling for the magnetic qBIC; SEM image of a ripped, rectangular metasurface; AFM retrace curves; raw data for resonance scaling in Figure 4c (PDF)

## AUTHOR INFORMATION

**Corresponding Author**

> **Andreas Tittl** - Chair in Hybrid Nanosystems, Nanoinstitute Munich, Faculty of Physics, Ludwig-Maximilians-Universität, Munich, Germany. Email: Andreas.Tittl@physik.uni-muenchen.de

**Authors**

> **Michael Hirler** - Chair in Hybrid Nanosystems, Nanoinstitute Munich, Faculty of Physics, Ludwig-Maximilians-Universität, Munich, Germany; https://orcid.org/0009-0008-4154-5791
>
> **Alexander A. Antonov** - Chair in Hybrid Nanosystems, Nanoinstitute Munich, Faculty of Physics, Ludwig-Maximilians-Universität, Munich, Germany
>
> **Enrico Baù** - Chair in Hybrid Nanosystems, Nanoinstitute Munich, Faculty of Physics, Ludwig-Maximilians-Universität, Munich, Germany
>
> **Andreas Aigner** - Chair in Hybrid Nanosystems, Nanoinstitute Munich, Faculty of Physics, Ludwig-Maximilians-Universität, Munich, Germany
>
> **Connor Heimig** - Chair in Hybrid Nanosystems, Nanoinstitute Munich, Faculty of Physics, Ludwig-Maximilians-Universität, Munich, Germany

**Haiyang Hu** – Chair in Hybrid Nanosystems, Nanoinstitute Munich, Faculty of Physics, Ludwig-Maximilians-Universität, Munich, Germany

**Author Contributions**

A.A. and M.H. developed the conceptual idea. A.A.A. and M.H. conducted simulations. M.H. performed fabrication, SEM and optical measurements with contributions from H.H.. E.B. carried out AFM measurements and corresponding data analysis. C.H. contributed to ellipsometry measurements. A.T., A.A. and A.A.A. supervised the research. The manuscript was written through contributions of all authors. All authors have given approval to the final version of the manuscript.

**Notes**

The authors declare no competing financial interest.

**ACKNOWLEDGMENT**

Funded by the European Union (EIC, OMICSENS, 101129734, EIC, NEHO, 101046329, ERC, METANEXT, 101078018). Views and opinions expressed are however those of the author(s) only and do not necessarily reflect those of the European Union or the European Research Council Executive Agency. Neither the European Union nor the granting authority can be held responsible for them. This project was also funded by the Deutsche Forschungsgemeinschaft (DFG, German Research Foundation) under grant numbers EXC 2089/1–390776260 (Germany's Excellence Strategy) and TI 1063/1 (Emmy Noether Program), the Bavarian program Solar Energies Go

Hybrid (SolTech) and the Center for NanoScience (CeNS). The authors thank Maxim Gorkunov for valuable discussions.

**Supporting Information for**

# Accessible, All-Polymer Metasurfaces: Low Effort, High Quality Factor

*Michael Hirler, Alexander A. Antonov, Enrico Baù, Andreas Aigner, Connor Heimig, Haiyang Hu and Andreas Tittl**

Chair in Hybrid Nanosystems, Nanoinstitute Munich, Faculty of Physics, Ludwig-Maximilians-Universität, Munich, Germany

* Email: Andreas.Tittl@physik.uni-muenchen.de

## Contents

**Supplementary Note 1 – Refractive Index Data of PMMA**

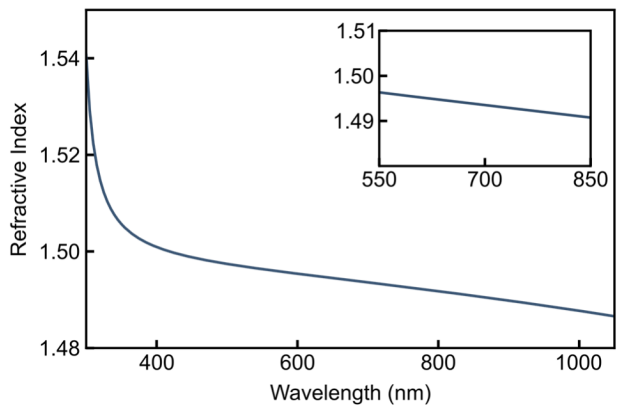

**Figure S1.** Real Part of the Refractive Index of PMMA as Determined in Experiment. In the respective wavelength region, losses are negligible. A 300 nm thick PMMA film (AR-P 679-04, Allresist) was spin-coated on a silicon substrate (3600 rpm for 1 min) and measured by white-light spectral ellipsometry.

**Supplementary Note 2 – Numerical Modelling for the qMagnetic BIC**

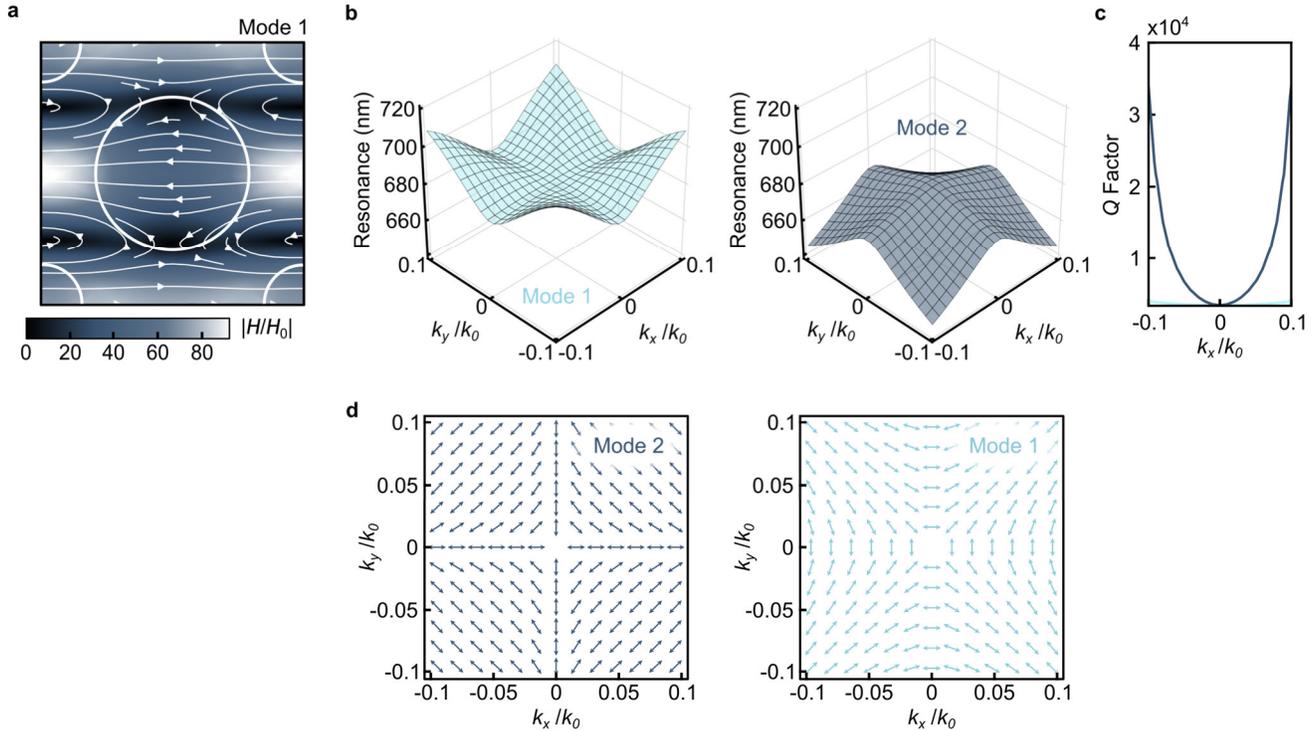

**Figure S2.** Numerical Modelling for the Magnetic BIC. (a) Magnetic near field enhancement $|H/H_0|$ as well as resonance position (b) and $Q$ factor (c) in momentum space for the magnetic mode in Figure 2b. (d) Far field polarizations of the two magnetic modes with topological charges -1.

**Supplementary Note 3 – SEM Image of Ripped, Rectangular Metasurface**

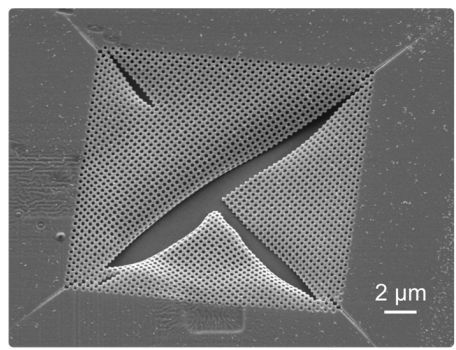

**Figure S3.** SEM Image of Ripped, Rectangular Metasurface. This exemplarily demonstrates the tendency of the PMMA membranes to rip at structural corners. The viewing angle is 25°.

## Supplementary Note 4 – AFM Retrace Curves

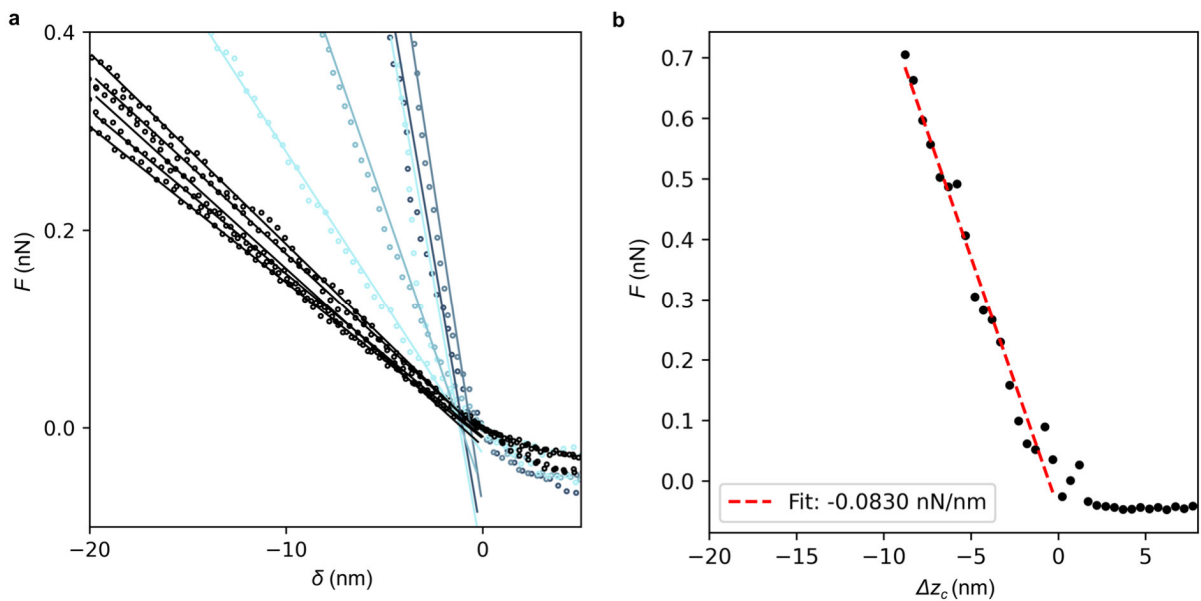

**Figure S4.** AFM Retrace Curves. Force-displacement ($F$-$\delta$) curves for (a) all points in Figure 3e and (b) silicon for determination of $\Delta z_c$. The lines denote fitting of Eq. (1).

## Supplementary Note 5 – Raw Data for Resonance Scaling

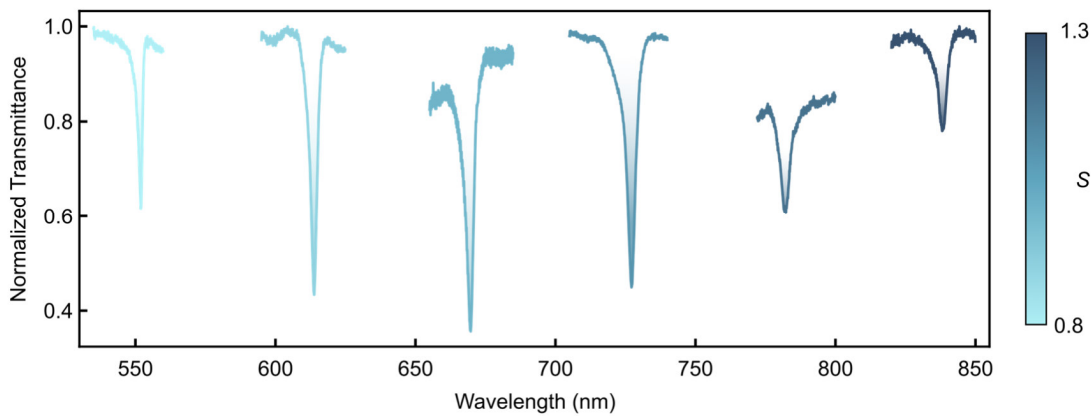

**Figure S5.** Raw Transmittance Spectra for Resonance Scaling without Normalization. The spectra correspond to membranes with $\alpha = 0.5$ and various lateral scaling factors $S$. In contrast, the spectra in Figure 4c in the main text were normalized such that the minimum and maximum transmittance equals 0 and 1, respectively.